\documentstyle[preprint,aps]{revtex}
\begin{document}
\draft

\author{A. R. Denton* and J. Hafner}

\address
{Institut f\"ur Theoretische Physik, Technische Universit\"at Wien\\
Wiedner Hauptstra{\ss}e 8-10, A-1040 Wien, Austria}

% force linebreaks with \\
\title{Thermodynamically Stable One-Component Quasicrystals:\\
A Density-Functional Survey of Relative Stabilities}

\date{\today}
\maketitle

\begin{abstract}
A combination of classical density-functional theory and
thermodynamic perturbation theory is applied to a survey of
finite-temperature trends in the relative stabilities of 
one-component crystals and quasicrystals 
interacting via effective metallic pair potentials 
derived from pseudopotential theory.  
Splitting the pair potential into a strongly repulsive short-range core
and a relatively weak oscillating tail, the Helmholtz free energy functional
of specified solid structures with Gaussian density distribution is computed 
for given thermodynamic states, and variationally minimized 
with respect to the width of the distribution.
Comparing the free energies of several periodic crystals and 
rational approximant models of quasicrystals over a range of 
pseudopotential parameters, {\it thermodynamically} stable quasicrystals
are predicted for parameters approaching the limits of mechanical stability 
of the crystal structures.  
Quasicrystalline stability is attributed to vibrational stiffness 
and energetically favorable medium- and long-range interactions.
The results support and significantly extend conclusions of 
previous ground-state lattice-sum studies.
\end{abstract}

\bigskip
\pacs{PACS numbers: 61.44.+p, 64.70.Dv, 61.25.Mv, 05.70.-a}

%\narrowtext
%\twocolumn

%%%%%%%%%%%%%%%%%%%%%%%%%%%%%%%%%%%%%%%%%%%%%%%%%%%%%%%%%%%%%%%%%%%%%%%%%%%
%%%%%%%                       BODY OF TEXT
%%%%%%%%%%%%%%%%%%%%%%%%%%%%%%%%%%%%%%%%%%%%%%%%%%%%%%%%%%%%%%%%%%%%%%%%%%%

\section{Introduction}

Since the landmark discovery~\cite{Shechtman} of 
long-range icosahedral quasiperiodic ordering in Al-Mn alloys, 
thermodynamically stable quasicrystals have been produced 
in a rich variety of systems~\cite{QC1,QC2,Tsai}.
The fact remains, however, that all known quasicrystals are alloys 
of at least two metallic elements.  
A fundamental question has thus naturally arisen:  can one-component 
quasicrystals be thermodynamically stable?
An icosahedral quasiperiodic phase having lower ground-state enthalpy 
than periodic crystals has in fact been predicted for an idealized system 
of particles interacting via a square-well pair potential~\cite{Jaric1}.
The stability range is confined, however, to a very narrow range of 
well widths and pressures and the physical relevance is unclear.
On a much broader scale, 
the question has been addressed by extensive lattice-sum potential energy 
calculations~\cite{Smith,Windisch} for a variety of crystals and 
quasicrystals interacting via effective metallic pair potentials 
generated by pseudopotential theory.  
The important conclusion of these studies is that energetically stable 
ground-state one-component quasicrystals are indeed possible, albeit 
within a restricted range of pair potential parameters having no
counterparts in the Periodic Table.  

Although very useful for identifying general trends in relative 
structural stabilities, the lattice-sum approach, because it takes 
into account only internal energy, is restricted to zero temperature.
Therefore it cannot directly address the general issue of 
{\it thermodynamic} stability.
An alternative theoretical approach, especially suited to finite temperatures, 
is the classical density-functional (DF) approach~\cite{DF1}, by which 
the thermodynamic free energy is determined as a variational minimum 
with respect to density of an approximate free energy functional.
Important fundamental applications have already been made to
systems of hard spheres with quasiperiodic order~\cite{MA,Jaric2}.
Distinguished as they are by purely entropic (structure-dependent) 
free energy, hard-sphere (HS) systems are ideally suited for focusing 
exclusively on the role of entropy in stabilizing competing structures.  
Because the infinitely repulsive HS pair potential lacks an energy scale, 
however, temperature in this case is irrelevant to static properties.
Comparing elementary crystals with model quasicrystalline structures, 
the quasicrystals were predicted to be either higher in free energy and thus
metastable (in fact, relative even to the fluid phase)~\cite{MA} or 
mechanically unstable (no variational minimum)~\cite{Jaric2}.
These studies clearly suggest that ordinary entropy 
(associated with phonons and disorder) is by itself 
insufficient to stabilize quasiperiodic ordering.  
The previous lattice-sum and DF studies leave open, however, the question 
of whether at finite temperatures longer-range interactions 
can conspire to stabilize one-component quasicrystalline structures.

Our main goal in this paper is to bridge the gap between previous studies 
by extending classical DF methods to simple metallic systems 
whose effective pair interactions may be derived from pseudopotential theory.
By varying the parameters that define the pair potential, we are able to
systematically survey a range of physically relevant systems and 
thermodynamic states for the possible existence of stable one-component 
quasicrystals.  
As a structural model, we focus here on a certain class of
rational approximants, periodic crystals with large unit cells 
whose atomic configurations realistically model quasicrystalline order.
As our main result, we indeed find, over a limited range of pseudopotential 
parameters, thermodynamically stable one-component quasicrystals, whose 
stabilities are largely attributable to medium- and long-range interactions.
As an incidental result, we also predict mechanically stable, but
thermodynamically unstable, HS rational approximants, thus supporting 
conclusions of previous DF studies for HS systems~\cite{MA,Jaric2}.

We proceed in Sec.~II by outlining the calculation of the effective metallic 
pair potentials.  These constitute a fundamental input to the theory
described in Sec.~III, which combines classical DF methods with thermodynamic 
perturbation theory.  
In Sec.~IV, after first describing the quasicrystal structural model, 
we present results, from a broad survey of pseudopotential parameters, 
for trends in the relative stabilities of various solid structures. 
Finally, in Sec.~V we summarize and conclude with some remarks on the 
implications of our study for the stabilities of real quasicrystals.

\section{Interatomic Interactions in Metals}

Interactions between ions in a metal are complicated by the presence
of conduction (or valence) electrons, whose response to the ions 
in many cases must be determined by elaborate methods of 
electronic DF theory~\cite{Jones}.
However, when the energy levels of core and conduction electrons are 
sufficiently well separated, as they are in the simple metals 
({\it e.g.}, Al, Mg, and the alkalis), the conduction electrons may be 
justly regarded as nearly free~\cite{AS,HM}.
It then proves possible to replace the strong electron-ion potential 
by a much weaker {\it pseudopotential}, which has no bound states 
but preserves the true valence energy spectrum.

The pseudopotential approach~\cite{Harrison,Heine,Hafner1} has been refined 
over three decades and, with remarkable success, has helped to explain 
a wide range of structural and thermodynamic properties of solids, including
variations in crystal structure through the Periodic Table~\cite{HH}.
Its conceptual appeal lies in its reduction of the two-component system
of electrons and ions to an effective one-component system of pseudoatoms.  
The reduction
is achieved by regarding the pseudopotential as a weak perturbation to
a uniform gas of valence electrons, which is further assumed to respond 
adiabatically and linearly to the effective static external potential 
imposed by the ions.  In linear response theory, the Fourier components
of the induced electron density $\hat\rho_e({\bf k})$ and the ion density
$\hat\rho_i({\bf k})$ are simply related by
\begin{equation}
\hat\rho_e({\bf k})=\chi(k) \hat{v}^{ps}(k)\hat\rho_i({\bf k}),
\label{linresp}
\end{equation}
where $\chi(k)$ is the static electron-density response function.
To second order in perturbation theory, the effective ion-ion pair potential 
then takes the Fourier-space form 
\begin{equation}
\hat\phi(k) = Z^2v(k)\left[1+\left(\frac{1}{\epsilon(k)}-1
\right)\left(\frac{{\hat{v}^{ps}}(k)}{Zv(k)}\right)^2\right],
\label{pp}
\end{equation}
where $v(k)=4{\pi}e^2/k^2$ is the Fourier transform of the bare Coulomb 
interaction and $\epsilon(k)$ is the dielectric function, whose 
density dependence naturally gives rise to {\it density-dependent} 
effective pair potentials.
The dielectric and static response functions are closely related through
\begin{equation}
\epsilon(k)=1-\left(\frac{v(k)\chi_o(k)}{1+v(k)G(k)\chi_o(k)}\right),
\label{eps}
\end{equation}
where $\chi_o(k)$ is the Lindhard free electron density response function
and $G(k)$ is the local field correction, which
incorporates exchange and correlation effects.  
Experience has shown~\cite{Hafner1} that the accuracy of the
effective pair potential is particularly sensitive to the choice of $G(k)$.  
Here we take the analytic form proposed by Ichimaru and Utsumi~\cite{IU},  
which accurately fits Monte Carlo data for the correlation energy and 
satisfies important self-consistency conditions.

The choice of pseudopotential is less critical, and in practice a 
parametrized form is often assumed and fitted to experimental data for 
transport and other properties sensitive to electron-ion interactions.  
A particularly simple and popular form, which we have adopted here, is 
the (local and energy-independent) empty-core pseudopotential~\cite{Ashcroft}
\begin{eqnarray}
v^{ps}(r)&=&0,\qquad\qquad r<r_c\nonumber\\
&=&-{{Ze^2}\over{r}},\qquad r>r_c,
\label{ps}
\end{eqnarray}
parametrized by valence $Z$ and core radius $r_c$. 
The resulting real-space effective pair potential $\phi(r)$ following from 
Eqs. (2-4) is illustrated in Fig.~1 for several values of
atomic volume $\Omega$ (or number density $1/\Omega$), valence, and core radius
as a ratio of the electron-sphere radius $r_s=(3{\Omega}/4{\pi}Z)^{1/3}$. 
(Note that we use atomic units throughout, representing energies in 
milli-Rydbergs and lengths in Bohr radii $a_o$.)  
As is typical for the simple metals, $\phi(r)$ is characterised by 
a steeply repulsive short-range core, arising from screening of the 
bare ion-ion interaction by polarization of the electrons, and 
a relatively weak oscillating tail, dominated by Friedel oscillations
with periodicity $\pi/k_F$ ($k_F$ being the Fermi wavevector magnitude).
Extensive analysis~\cite{Hafner1,HH} has shown that the form of the potential 
in the vicinity of the nearest-neighbor distance depends sensitively 
on the interplay between the core and tail regions.
In particular, with increasing electron density 
(increasing $Z$ or decreasing $\Omega$) at fixed $r_c/r_s$, 
the core shifts to shorter distances and the wavelength of the
Friedel oscillations shortens [Fig.~1 (a) and (b)].
Correspondingly, the core repulsion steepens 
and the depth of the first minimum becomes shallower as the oscillations
move under the core.  For sufficiently high electron density, the first 
minimum may be covered by the core, resulting in a repulsive shoulder, 
as seen in Fig.~1 (a) for the case of $Z=3.2$.  
On the other hand, with decreasing core radius for fixed $Z$, 
the Friedel oscillations weaken in amplitude and shorten in wavelength.  
For sufficiently small $r_c$, the first minimum similarly may be pushed 
to larger distances, as seen in Fig.~1 (c) for the case $r_c/r_s=0.4$.
In Sec.~IV B we discuss the influence of these properties of $\phi(r)$
on structural stabilities of solids.

\section{Theoretical Approach}

\subsection{Classical Density-Functional Theory}

With effective pair potentials in hand, we may now in a sense regard
the electronic structure problem as solved -- albeit trivially in this
case -- and henceforth consider, at fixed average density, a
purely classical one-component system of pair-wise interacting pseudoatoms.
We thus proceed to take $\phi(r)$ as input to a classical DF theory for 
the Helmholtz free energy, the central quantity of interest in assessing 
relative stabilities of competing solid structures.
The DF approach~\cite{DF1} provides a general framework 
for describing nonuniform systems, of which solids are extreme examples.  
For classical systems, 
it is founded on the definition of a functional $F[\rho]$ of the 
spatially-varying one-particle number density $\rho({\bf r})$, which 
satisfies a fundamental variational principle~\cite{Mermin,Evans},  
according to which $F[\rho]$ 
is minimized (at constant average density) by the equilibrium density, its
minimum value equaling the Helmholtz free energy.
For convenience, $F[\rho]$ is separated into ideal-gas and excess 
contributions:
\begin{equation}
F[\rho]=F_{id}[\rho]+F_{ex}[\rho].
\label{Fidex}
\end{equation}
The first term is the free energy of the nonuniform system 
in the absence of interactions, directly related to the 
configurational entropy, and is given exactly by
\begin{equation}
F_{id}[\rho]= k_BT\int d{\bf r}\rho({\bf r})[\ln(\rho({\bf r})\lambda^3)-1],
\label{Fid1}
\end{equation}
where $\lambda$ is the thermal de Broglie wavelength and $T$ is temperature.
The excess contribution depends entirely upon internal interactions and 
is not known exactly.
Approximations for $F_{ex}[\rho]$ abound, but experience has proven the 
class of so-called weighted-density approximations~\cite{Tar,CA1,CA2} (WDA)
to be especially useful in a variety of contexts~\cite{DF2}.
These are all based on a mapping of the nonuniform excess free energy onto 
its uniform counterpart, as discussed further below.
It should be emphasized that here we entirely ignore a
sizable volume-dependent, but structure-independent, term in the 
free energy associated with the electrons~\cite{AS,HM}.
This is justified by the fact that here
we are concerned only with the relative stabilities of different 
solid structures and thus always compare free energies of solids 
at the {\it same average density}.
In applications to phase transitions, however, the volume term would 
have to be included for coexisting phases having different densities.

Although there have been numerous successful applications of DF theory to
simple atomic systems~\cite{DF2} ({\it e.g.,} HS and Lennard-Jones 
potentials), few thus far have considered metals~\cite{Oxtoby,IKH}.
Studies with other long-range pair potentials have shown~\cite{LK,KB1},  
however, that weighted-density methods are often 
best applied in combination with thermodynamic perturbation theory.  
The essential reason is that a single mapping proves insufficient 
to accurately approximate both the excess entropy and the internal 
energy~\cite{CA2,Hasegawa}.
As seen from Fig.~1, the basic form of $\phi(r)$ for the simple metals
lends itself naturally to a perturbation approach, wherein the full pair 
potential is sensibly split into a short-range reference potential 
$\phi_o(r)$ and a perturbation potential 
$\phi_p(r)=\phi(r)-\phi_o(r)$.
Such an approach has already been applied successfully to 
the structure and thermodynamics of Lennard-Jones liquids~\cite{HM,WCA}
and solids~\cite{Tar,CA2,JA,KB2,Mederos}
and of metallic liquids~\cite{Hafner2,HK}
and solids~\cite{DHK}.
The Helmholtz free energy may be correspondingly separated and 
formally expressed in the exact form~\cite{Evans}
\begin{equation}
F[\rho] = F_o[\rho] + {1\over{2}}\int_0^1 d\lambda\int d{\bf r}\int d{\bf r}'
\rho^{(2)}({\bf r},{\bf r}';\phi_{\lambda})\phi_p(|{\bf r}-{\bf r}'|),
\label{pert1}
\end{equation}
where $\rho^{(2)}({\bf r},{\bf r}';\phi_{\lambda})$ is the two-particle
density for a system interacting via the pair potential
$\phi_{\lambda}(r)\equiv \phi_o(r)+\lambda\phi_p(r)$.
The reference free energy $F_o[\rho]$ includes the ideal-gas free energy
and the part of the excess free energy associated with short-range 
interactions, while the perturbation term is the remaining part of
the excess free energy deriving from medium- and long-range interactions.
Further expanding $\rho^{(2)}$ about the reference pair potential, and
retaining only the leading term, leads directly to the first-order 
perturbation approximation, 
\begin{equation}
F[\rho] \simeq F_o[\rho] + {1\over{2}}\int d{\bf r}\int d{\bf r}'
\rho({\bf r})\rho({\bf r}')g_o({\bf r},{\bf r}')
\phi_p(|{\bf r}-{\bf r}'|),
\label{pert2}
\end{equation}
where 
$g_o({\bf r},{\bf r}')$ is the pair distribution function of the 
reference system (radial distribution function for a uniform liquid).
%######################################################################
% Inserted Text:
%
The second-order term in the perturbation expansion $F_2$ may be expressed
in terms of the mean-square fluctuation of the total perturbation energy
\begin{equation}
\Phi_p\equiv\sum_{i<j}\phi_p(|{\bf r}_i-{\bf r}_j|)
\label{Phip}
\end{equation}
according to 
\begin{equation}
F_2=-{1\over{2k_BT}}\langle(\Phi_p-\langle\Phi_p\rangle_o)^2
\rangle_o,
\label{pert3}
\end{equation}
where $\langle\cdots\rangle_o$ denotes an ensemble average 
for the reference system.
In general, the $n$th term in the expansion $F_n$ may be expressed
in terms of the mean fluctuations
$\langle (\Phi_p-\langle\Phi_p\rangle_o)^m\rangle_o$
divided by $(k_BT)^{n-1}$, with $m\leq n$~\cite{HM,Zwanzig}.
Thus, roughly speaking, convergence of the expansion is determined by 
the magnitude of fluctuations in $\Phi_p$ relative to $k_BT$.
We return to the issue of convergence in Sec. V.
%
%######################################################################

Some freedom remains in the separation of the pair potential and in 
the choice of reference system.  Here we use the standard 
Weeks-Chandler-Andersen~\cite{WCA} (WCA) separation, which splits 
the potential at its first minimum and shifts it such that
\begin{eqnarray}
\phi_o(r)&=&\phi(r)-\phi(r_{min}),\qquad r<r_{min}\nonumber\\
 &=&0,\qquad\qquad\qquad\qquad r>r_{min}
\label{WCA1}
\end{eqnarray}
and 
\begin{eqnarray}
\phi_p(r)&=&\phi(r_{min}),\qquad r<r_{min}\nonumber\\
 &=&\phi(r),\qquad\quad r>r_{min}.
\label{WCA2}
\end{eqnarray}
The strongly repulsive reference system we map onto a system of hard spheres 
of effective diameter $d$.  Although the WCA prescription~\cite{HM,WCA} 
for $d$ is usually considered superior for liquids, application to solids 
is problematic because the required function $g_o({\bf r},{\bf r}')$ 
is relatively poorly known (see below).
Thus, we use instead the simpler Barker-Henderson prescription,
\begin{equation}
d = \int_0^{\infty}dr\{1-\exp[-\phi_o(r)/k_BT]\},
\label{BH}
\end{equation}
which for the dense liquid gives an effective diameter very close to that 
of the WCA prescription and in fact reduces to the latter as the steepness 
of $\phi_o(r)$ increases.

The reference free energy of the solid is now well approximated by the 
modified weighted-density approximation (MWDA), which is known to give 
an accurate description of HS systems~\cite{MWDA}.
This maps the excess free energy per particle of the HS solid 
onto that of the corresponding {\it uniform} fluid $f_o$, according to
\begin{equation}
F_{ex}^{MWDA}[\rho]/N = f_o(\hat\rho),
\label{MWDA}
\end{equation}
where the weighted density
\begin{equation}
\hat\rho \equiv \frac{1}{N} \int d{\bf r} \int d{\bf r}'
\rho({\bf r})\rho({\bf r}')w(|{\bf r}-{\bf r}'|;\hat\rho)
\label{rhohat1}
\end{equation}
is a self-consistently determined weighted average of the physical density.  
The associated weight function $w$ is in turn specified by the normalization 
condition
\begin{equation}
\int d{\bf r}' w(|{\bf r}-{\bf r}'|) = 1,
\label{norm}
\end{equation}
and by the requirement that $F_{ex}^{MWDA}[\rho]$ generate the 
exact two-particle (Ornstein-Zernike) direct correlation function 
$c(r)$ in the uniform limit:
\begin{equation}
\lim_{\rho({\bf r}) \to \rho} \left({{\delta^2 F_{ex}[\rho]}\over
{\delta\rho({\bf r})\delta\rho({\bf r}')}}\right)
= -k_BT c(|{\bf r}-{\bf r}'|;\rho).
\label{wf}
\end{equation}
Equations (\ref{norm}) and (\ref{wf}) lead to a simple analytic 
relation~\cite{MWDA} 
between $w(r)$ and the fluid-state functions $f_o$ and $c(r)$, computed here
using the convenient expressions following from the analytic solution of 
the Percus-Yevick integral equation for hard spheres~\cite{HM}, which are 
sufficiently accurate when evaluated at the relatively low weighted densities 
of relevance. 

In principle, the perturbation free energy $F_p$ 
[second term in Eq.~(\ref{pert2})] 
demands detailed knowledge of the HS solid pair distribution function 
$g_{HS}({\bf r},{\bf r}';d)$.  Although its translational average
is partially known from Monte Carlo studies~\cite{MC},  
the published data and analytic fits 
are limited to particular solid structures (close-packed crystals) 
at specific densities.
In many applications, however, the function is needed over a range of 
densities, and here also for a variety of solid structures.  For practical 
purposes, various approximations for $F_p$ have been proposed.  
Jones and Ashcroft~\cite{JA} and Curtin~\cite{Curtin} have calculated the
free energy of the LJ solid by 
approximating the spherical average of the two-particle 
density in the solid by that in the isotropic liquid, 
thereby ignoring any dependence upon solid structure.
Clearly this is insufficient for comparing
relative stabilites of different solid structures.  
Curtin and Ashcroft~\cite{CA2} have derived a structure-dependent correction, 
involving the difference between the true and reference system 
direct correlation functions, and have employed an empirical prescription
to study the Lennard-Jones crystal.
More recently, Mederos {\it et al.}~\cite{Mederos} have proposed, in the 
spirit of the WDA, mapping the pair distribution function of the solid onto 
the corresponding radial distribution function of the fluid, but evaluated
at an effective density chosen to satisfy a local compressibility relation.  
In practice, the requisite density is found to be so low that
the approximation is essentially equivalent to replacing 
$g_{HS}({\bf r},{\bf r}';d)$ by its ideal-gas limit, the 
Heaviside unit step function $u(|{\bf r}-{\bf r}'|-d)$, defined by
\begin{eqnarray}
u(x)&=&0,\qquad x<0\nonumber\\
 &=&1,\qquad x>0.
\label{u}
\end{eqnarray}
This amounts to a mean-field approximation in which self-correlations
are forbidden but correlations are otherwise entirely neglected.
The underlying justification~\cite{CA1} is the fact that in an ordered solid, 
unlike in a dense liquid, most of the structure of the two-particle density 
is contained already at the level of the one-particle density, rendering
$g({\bf r},{\bf r}')$ a relatively structureless function.
Because of its practicality and its success when applied to the 
Lennard-Jones~\cite{Mederos} and other~\cite{Likos,DL} systems, we have 
adopted this approximation for the perturbation free energy. 

Finally, collecting together the above-mentioned DF and perturbation 
approximations, our working free energy functional is expressed as
\begin{equation}
F[\rho] \simeq F_{HS}^{MWDA}[\rho;d] + 
{1\over{2}}\int d{\bf r}\int d{\bf r}'
\rho({\bf r})\rho({\bf r}')u(|{\bf r}-{\bf r}'|-d)
\phi_p(|{\bf r}-{\bf r}'|),
\label{F}
\end{equation}
which is to be variationally minimized
with respect to the solid number density $\rho({\bf r})$.
%########################################################################
% Modified Text:
%
The perturbation free energy evidently amounts to a lattice-sum over 
$\phi_p(r)$, but one that incorporates thermal broadening of the 
density distribution.
Note that the temperature dependence of the total free energy consists
of a trivial (linear) dependence of the HS reference free energy
and a non-trivial dependence through the effective HS diameter $d$ 
[Eq.~(\ref{BH})].
%
%########################################################################

The free energy of the liquid phase is calculated, at the same level of
perturbation theory, via the uniform limit of Eq.~(\ref{pert2}).  
Here the reference system free energy is very accurately approximated by
the Carnahan-Starling excess free energy of the HS fluid~\cite{HM}.
For the perturbation, however, a mean field approximation is inappropriate, 
since the radial distribution function of the dense HS fluid is far more 
structured than that of the solid.
Therefore, instead of Eq.~(\ref{u}), we use the highly accurate 
Verlet-Weis fit to simulation data for the fluid function $g_{HS}(r)$.

\subsection{Practical Implementation}

Calculation of the free energy based on the approximate functional (\ref{F}) 
requires, in practice, specification of the solid structure, 
{\it i.e.}, the position vectors ${\bf R}$ of the equilibrium atomic sites.
Here we consider three elementary crystal structures (fcc, hcp, and bcc) 
and, as a model of quasicrystalline structure, a sequence of 
rational approximants~\cite{QC1,Goldman}.
The fcc and bcc crystals are Bravais lattices, characterized -- 
in fact, defined -- by equivalence of every lattice site~\cite{AM}. 
The hcp structure, is composed of two interpenetrating simple-haxagonal 
Bravais lattices or, equivalently, one simple-hexagonal lattice 
with a two-atom basis.
The rational approximants are experimentally observed periodic crystals
with large unit cells, whose structures realistically model the 
atomic ordering of certain known quasicrystals.  Their structures are
described in detail in Sec.~IV A.
For practical purposes, here we regard them as consisting of a 
simple-cubic Bravais lattice together with a multi-atom basis.  

In addition to the positions of the atomic sites, the density distribution 
about the sites must in practice be specified.
A particularly simple and widely adopted ansatz for the density 
distribution~\cite{Tar,DF2} places normalized isotropic Gaussians 
at the sites of the solid, according to
\begin{equation}
\rho({\bf r})=\left(\frac{\alpha}{\pi}\right)^{3/2}\sum_{\bf R}\sum_{\bf d}
\exp(-\alpha|{\bf r}-{\bf R}-{\bf d}|^2).
\label{Gauss}
\end{equation}
Here the first sum is over the sites (or unit cells) of the 
underlying Bravais lattice and the second is over the basis atoms, whose 
positions are described by displacements ${\bf d}$ relative to the
lattice sites.  The parameter $\alpha$ determines the width of the
distribution and represents the single variational parameter with
respect to which the free energy functional is minimized.
The Gaussian ansatz has been shown by simulation~\cite{YA,OLW}
and by numerical analysis~\cite{LMH} to yield a reasonably accurate
approximation to the true density distribution in close-packed 
Bravais crystals near melting, despite deviations in the detailed form
of the distribution, and it has been widely and successfully used in 
diverse applications of classical DF theory to thermodynamic properties 
of crystals~\cite{DF2}.
Implicit in Eq.~(\ref{Gauss}) is the assumption that the distribution 
is identical about every site of the solid.  While rigorously true 
of the Bravais lattices, this assumption is not strictly valid for 
the other structures considered.  
Although, in principle, a different variational width parameter could be 
assigned to each distinct site, or even free minimization could be 
performed~\cite{OLW}, in practice the associated numerical cost of 
multi-dimensional minimization would make prohibitive such an 
extensive survey of pair potentials, solid structures, and 
thermodynamic states as we undertake here.  
In any case, excepting the most open structures, 
the density distribution is expected to be isotropic and Gaussian 
to within a reasonable approximation, as discussed in Sec.~V.
Thus, all of our free energy calculations make use of 
the Gaussian ansatz (\ref{Gauss}).

The density of any periodic solid may be represented as a Fourier series
of the general form
\begin{equation}
\rho({\bf r})=\frac{1}{V}\sum_{\bf G}\rho_{\bf G}
{\rm e}^{i{\bf G}\cdot{\bf r}},
\label{rhor}
\end{equation}
where $V$ is the total volume and the sum is over the reciprocal lattice 
vectors (RLV) ${\bf G}$ of the underlying Bravais lattice. 
For the density defined by Eq.~(\ref{Gauss}), the Fourier transform of 
$\rho({\bf r})$ may be factorized as
\begin{equation}
\rho_{\bf G}\equiv \int d{\bf r} {\rm e}^{-i{\bf G}\cdot{\bf r}}\rho({\bf r}) 
= N\exp(-G^2/4\alpha)S_{\bf G},
\label{rhoG}
\end{equation}
where 
\begin{equation}
S_{\bf G}=\frac{1}{N_b}\sum_{\bf d}{\rm e}^{-i{\bf G}\cdot{\bf d}}
\label{SG}
\end{equation}
is a geometrical structure factor and $N_b$ is the number of basis atoms.
Substituting Eqs.~(\ref{rhor}) and (\ref{rhoG}) into Eq.~(\ref{rhohat1}), 
the weighted density in the MWDA may then be expressed in the form
\begin{equation}
\hat\rho = \rho_s\left[1+\sum_{G\neq 0}\exp(-G^2/2\alpha) w_G 
\sum_{\rm deg}|S_{\bf G}|^2\right],
\label{rhohat2}
\end{equation}
with $\rho_s\equiv N/V$ being the average solid density, $w_G$ the 
Fourier transform of the weight function, and where we have used the property 
$S_{-{\bf G}}=S^*_{\bf G}$.  The first sum in Eq.~(\ref{rhohat2}) is over 
magnitudes of the RLV's, and the second is over degeneracies, 
that is, over all RLV's having the same magnitude $G$.
Note that for a given structure, the latter sum need be computed only once 
for each magnitude $G$ (most efficiently in complex arithmetic).
Computation of $\hat\rho$ via Eq.~(\ref{rhohat2}) is now achieved by a 
straightforward summation over RLV magnitudes, and substitution into 
Eq.~(\ref{MWDA}) then immediately gives the HS reference excess free energy 
[first term on right side of Eq.~(\ref{F})].

For density distributions sufficiently narrow that neighboring Gaussians 
do not significantly overlap, the ideal-gas free energy per particle 
[Eq.~(\ref{Fid1})] is very accurately approximated by 
\begin{equation}
F_{id}/N = {3\over{2}}k_BT{\ln}(\alpha\lambda^2),
\label{Fid2}
\end{equation}
to within an irrelevant additive constant.
Typically, Eq.~(\ref{Fid2}) is valid for ${\alpha}d^2>50$, a condition 
always satisfied at the densities of interest here.

The perturbation free energy [second term on right side of Eq.~(\ref{F})] 
is similarly computed in Fourier space by first rewriting it in the form
\begin{equation}
F_p[\rho;d]=\frac{1}{2}\int d{\bf r} \int d{\bf r}' \rho({\bf r})\rho({\bf r}')
[\phi_p(|{\bf r}-{\bf r}'|)+\phi_p(d)\theta(|{\bf r}-{\bf r}'|-d)],
\label{Fp1}
\end{equation}
where $\theta$ denotes the Mayer function for hard spheres, defined by
$\theta(x)\equiv u(x)-1$.
The Fourier space representation is then easily shown to be
\begin{equation}
F_p[\rho;d]/N=\frac{1}{2}\rho_s\sum_G\exp(-G^2/2\alpha)
\left[\tilde\phi_p(G)+\phi_p(d){\tilde\theta}(G)\right]
\sum_{\rm deg}|S_{\bf G}|^2,
\label{Fp2}
\end{equation}
where
\begin{equation}
\tilde\phi_p(G)\equiv\int d{\bf r} {\rm e}^{-i{\bf G}\cdot{\bf r}}\phi_p(r)
\label{phiG}
\end{equation}
is the Fourier transform of the perturbation potential, and 
\begin{equation}
{\tilde\theta}(G)\equiv\int d{\bf r} {\rm e}^{-i{\bf G}\cdot{\bf r}}\theta(r-d) 
= -\frac{4\pi}{G^3}\left[\sin(Gd)-Gd\cos(Gd)\right].
\label{thetaG}
\end{equation}
In practice, $\tilde\phi_p(G)$ is readily computed by a 
fast Fourier transform algorithm.
Care must be taken to sum Eqs.~(\ref{rhohat2}) and (\ref{Fp2}) 
over a sufficiently high number of RLV's to ensure covergence.

For fixed $\rho_s$, $T$, and pseudopotential parameters $Z$ and $r_c$, 
the free energy functional is computed from Eqs.~(\ref{MWDA}), (\ref{F}), 
(\ref{rhohat2}), (\ref{Fid2}), and (\ref{Fp2}) for a given solid structure 
and then minimized with respect to $\alpha$ to obtain the free energy.
For the hcp crystal, additional minimization with respect to
the $c/a$ ratio is performed.
The existence of a variational minimum implies that the chosen solid is 
at least mechanically stable.  Finally, comparing the free energies of
different solid structures and of the liquid, the thermodynamically stable 
structure is determined as that having the lowest free energy.

\section{Survey of Relative Stabilities}

\subsection{Quasicrystal Structural Model}

The quasicrystalline structure is modelled here by a three-dimensional
Penrose tiling constructed by the projection of a six-dimensional
hypercubic lattice onto the three-dimensional physical space, such that
the six hypercubic basis vectors are projected onto six vectors
forming an icosahedral basis 
\begin{equation}
\{{\bf e}_i; i=1,\ldots,6\}=
c\Bigl\{(1,\tau,0) + {\rm cyclic~permutations~(c.p.)};
(-1,\tau,0) + {\rm c.p.}\Bigr\},
\label{epar}
\end{equation}
with $c=(1+\tau^2)^{-1/2}=(2+\tau)^{-1/2}$, where $\tau=(1+\sqrt{5})/2$
is the golden mean.  To each of the basis vectors ${\bf e}_i$ there corresponds
in the three-dimensional space orthogonal to the physical space a basis vector
\begin{equation}
\{{\bf e}_i^{\perp}; i=1,\ldots,6\}=
c\Bigl\{(-\tau,1,0) + {\rm c.p.};
(\tau,1,0) + {\rm c.p.}\Bigr\}.
\label{eperp}
\end{equation}
The Penrose lattice consists of all points whose images in six dimensions
lie inside an acceptance domain $V_A^{\perp}$ with the form of a 
rhombic triacontahedron described by 
\begin{equation}
V_A^{\perp}\equiv\{{\bf x}_o^{\perp}+\sum_{i=1}^6\xi_i\cdot{\bf e}_i^{\perp};
0\leq\xi_i\leq 1\}.
\label{domain}
\end{equation}
The basic units of the Penrose lattice are prolate and oblate rhombohedra
spanned by the basis vectors $({\bf e}_1,{\bf e}_5,{\bf e}_6)$ and
$({\bf e}_2,{\bf e}_5,{\bf e}_6)$, respectively.  If all vertices of a
Penrose lattice are occupied by atoms, three types of bonds can be 
distinguished: $a$-bonds along the edge of the rhombohedra, $b$-bonds of 
length $1.05 a$ along the short diagonal of a rhombic face, and $c$-bonds 
of length $0.56 a$ along the short body diagonal of the oblate rhombohedra.
Note that second-nearest neighbors connected via two $c$-bonds appear 
at a distance of $1.12 a$.  The existence of the short $c$-bonds limits
the diameters of the atoms that can be placed on the vertices and leads
to a very low packing fraction.  In the ``unit-sphere packing" on a
Penrose lattice proposed by Henley~\cite{Henley}, a more compact structure
is achieved by occupying only one of the sites on both ends of a $c$-bond.
Since there is no essential asymmetry between these two sites, the choice
of the empty sites is subject to some degree of arbitrariness.  The choice
is unique only in the case of $c$-bonds forming chains with an even number
of members: in this case the end-points and every second point along the
chain are occupied.  Except for a small number of end-points of $c$-chains, 
the empty sites have four $a$-, five $b$-, and two $c$-bonds, and are
surrounded by two oblate and two prolate rhombohedra forming together
a rhombic dodecahedron.  From the frequency of the dodecahedra, the 
average occupancy of the vertices of the Penrose lattice can be estimated
as $0.7802$ (for further details, see Henley~\cite{Henley}), 
leading to a maximum packing fraction for the hard-sphere quasicrystal of 
$\eta_{QC}=0.6288$, rather close to the highest achievable packing 
fraction for a random packing of hard spheres, $\eta_{RPHS}=0.636$.
(The maximum packing fraction, defined as the volume ratio, with respect to
the total volume, of tightly packed hard spheres, provides a 
standard measure of packing efficiency.)

The elimination of a specific class of lattice sites from the Penrose tiling
creates a certain complication: if the images of the empty sites are
eliminated from the acceptance domain $V_A^{\perp}$, a non-compact
multiply-connected domain is formed.  As a consequence, techniques for
performing real and reciprocal quasi-lattice sums based on integrations
over the acceptance domain in perpendicular space are not easily applicable
to the unit-sphere packing.  A way out of the dilemma is to perform the
calculations for a hierarchy of rational approximants coming arbitrarily
close to the quasiperiodic limit.  A rational approximant is a periodic
structure that is obtained if, in the icosahedral basis in perpendicular 
space $\{{\bf e}_i^{\perp}\}$, the golden mean $\tau$ is replaced by a
rational number $\tau_n=F_{n+1}/F_n$, where $F_n$ is a Fibonacci number
defined by the recursive relation $F_{n+1}=F_n+F_{n-1}$, with $F_0=F_1=1$.
As an approximant is simply a crystal with a large unit cell, 
conventional techniques for performing lattice sums can be applied.  
Table~I lists the number of basis atoms $N_b$ in the periodically repeated 
unit cell and the maximum packing fraction $\eta_m$ for the first four 
generations of approximant to the quasiperiodic unit-sphere packing.  
Note that for the fourth-order approximant $\eta_m$ is already very close 
to the quasiperiodic limit, and that all of the approximants 
have maximum packing fractions lower than those of the crystals examined.
For a more detailed discussion of rational approximants to the 
Penrose lattice see, {\it e.g.}, Kraj\^c\'i and Hafner~\cite{KH}.

For comparison with previous studies, we note that the lattice-sum 
calculations of Smith~\cite{Smith} were performed on quasicrystal structures 
with various compact non-convex acceptance domains 
selected so as to maximize the packing fraction.
The most compact packing is achieved for an acceptance domain in the
form of a ``truncated stellated dodecahedron" (TSD) ($\eta_{TSD}=0.628$).
Only this type of quasicrystal was found to be more stable than
the tested crystalline structures in a certain range of parameter space
(valence, pseudopotential core radius, and atomic volume).
The DF study of M$^c$Carley and Ashcroft~\cite{MA}, for a system of 
hard spheres, also found the TSD acceptance domain to produce the 
most stable quasicrystal, although for that system only metastable.
The lattice-sum calculations of Windisch~\cite{Windisch}, on the other 
hand, dealt directly with the same rational approximants to the 
quasicrystalline unit-sphere packing that we examine here by DF theory.

\subsection{Results and Discussion}

As a simple but essential test of the theory, we first apply it to the 
hard-sphere reference system alone.  Figure~2 shows the predicted free energy 
per volume (in thermal units) as a function of average density for the various 
solid structures discussed above, as well as for the HS fluid.  (Note that, 
because of the infinitely steep repulsion of the HS potential at contact, 
the free energy has only a trivial linear temperature dependence, 
{\it i.e.}, $F/k_BT$ is independent of $T$.)
At lower densities ($\rho\sigma^3<0.91$) the fluid phase is 
thermodynamically stable, while at higher densities 
($\rho\sigma^3>1.04$) 
the close-packed crystals become inevitably stable.  
The fcc and hcp crystals are found to be essentially degenerate -- within the
presumed accuracy of the theory -- consistent with simulation results, the
minimizing hcp $c/a$ ratio increasing gradually over the displayed density 
range from $1.633$ (very close to the ideal value of $\sqrt{8/3}$) to $1.673$.  
The bcc crystal -- constrained by the density parametrization to be 
stable against shear -- becomes more stable than the fluid at higher densities, 
but remains metastable relative to the close-packed crystals.
Interestingly, the rational approximants are mechanically stable over a 
comparatively wide range of densities, including lower densities 
where the crystals are unstable.  
The essential reason is that their density distributions are 
narrower than those of the crystals at the same average density (see below).
Evidently, however, they remain metastable relative to both the fluid and 
the crystals.  Furthermore, at a given density, their free energies decrease 
in order of increasing maximum packing fraction (cf. Table~I), clearly 
demonstrating that the purely entropic HS system overwhelmingly favors 
packing efficiency.
It is interesting to compare these results with those from a similar
calculation, based on the same functional, for a HS glass~\cite{Loewen}.
For the random close-packed (RCP) model, the glass was found to be 
always metastable with respect to the fluid for $\eta_{RCP}<0.672$, 
which is significantly above the maximum approximant packing fraction.

As a further test, we have applied the full perturbation theory to a system
of rare gas atoms interacting via the Lennard-Jones (LJ) pair potential, 
a standard model that includes both short-range repulsive and long-range 
attractive interactions.  
In comparison with the results for the HS system, the qualitative picture 
is unchanged.  The theory predicts the close-packed crystals to form the 
stable high-density phase, in agreement with simulations, and the 
rational approximants to be always at best metastable.  
The clear implication is that long-range attractive interactions 
are not, by themselves, sufficient to stabilize quasicrystalline structures.

As a final test of the theory, as well as of the pseudopotential approach
outlined in Sec.~II, we have examined the freezing transitions of two 
simple metallic elements, Mg ($Z=2$, $r_c=1.31 a_o$) 
and Al ($Z=3$, $r_c=1.11 a_o$).
For both metals, the theory predicts coexisting liquid and solid densities
and structural energy differences in reasonable agreement with experiment.  
Furthermore, for Mg it predicts hcp 
to be the stable equilibrium structure, while for Al it predicts fcc, 
consistent with observation.  Details of this study will be presented 
elsewhere~\cite{DHK}.

With reasonable grounds for confidence in the predictive power of the theory, 
we now proceed to a broad (and more ambitious) survey of the simple metals.
Tables~II-IV summarize our results over a range of pseudopotential parameters, 
$2.0\leq Z\leq 3.4$ and $0.400\leq r_c/r_s\leq 0.575$.
Note that although $Z$ is of course an integer for any of the elements, 
we treat it here as a continuous parameter so as to better study trends 
in relative stabilities.
The thermodynamic states represented are defined by fixed temperature $T=500 K$ 
and atomic volume (or average solid density) 
${\Omega}=120 a_o^3$ ($\rho_s=0.0562 \AA^{-3}$), 
${\Omega}=150 a_o^3$ ($\rho_s=0.0450 \AA^{-3}$), and
${\Omega}=180 a_o^3$ ($\rho_s=0.0375 \AA^{-3}$), for Tables~II-IV, 
respectively.
Tabulated for each parameter set are, from top to bottom:
({\it i}) the most stable structure, of the liquid and the seven solids tested, 
determined as that having the lowest free energy
(structures in parentheses being metastable with respect to the liquid); 
({\it ii}) the corresponding Lindemann ratio $L$, defined as the ratio of 
the root-mean-square atomic displacement (away from the equilibrium site) 
to the nearest-neighbor distance, and given 
in the Gaussian approximation [Eq.~(\ref{Gauss})] by
\begin{equation}
L = \left(\frac{3}{2\alpha}\right)^{1/2}/
\left(\frac{6}{\pi}{\Omega}\eta_m\right)^{1/3},
\label{L}
\end{equation}
$\alpha$ being the minimizing Gaussian width parameter; and
({\it iii}) the effective HS diameter $d$, computed from Eq.~(\ref{BH}).

Tables~II-IV are dense with information.
For purposes of orientation, we note that hcp-Mg occurs at 
$r_c/r_s=0.539$, $0.501$, and $0.471$, corresponding to compressed, 
near-equilibrium, and expanded states, respectively. 
Similarly, expanded fcc-Al would occur at $r_c/r_s=0.523$, $0.486$, 
and $0.457$, were it not that at such low densities the fcc crystal 
is unstable (see below).
Physical interpretation is greatly aided by considering the effective 
HS packing fraction, defined by $\eta\equiv \frac{\pi}{6}\rho d^3$.
By default, the liquid is considered the stable phase when $\eta$ is so low 
that none of the solid structures is stable against thermally induced 
atomic displacements.  (The states considered are well removed 
from the vapour phase.)
At sufficiently high $\eta$, the close-packed fcc and hcp crystals 
tend to be thermodynamically stable, the minimizing hcp $c/a$ ratio 
lying close to, although usually slightly above, the ideal ratio.
This is to be expected, since experience with the HS system indicates
that strongly repulsive short-range interactions 
heavily favor packing efficiency.
In contrast to the HS and LJ systems, the fcc and hcp free energies are 
generally sufficiently different to clearly distinguish their
relative stabilities, reflecting some influence of longer-range interactions.
The more open bcc structure is always at best metastable, and thus appears 
nowhere in the stability tables.
In these qualitative respects, the simple metals do not differ significantly
from the HS and LJ systems.
Remarkably, however, in a relatively narrow band of parameter space, 
corresponding to intermediate effective packings, 
the rational approximants are predicted to be metastable or, in some cases, 
even {\it thermodynamically} stable. 
%########################################################################
% Inserted Text:
%
Although this parameter range contains none of 
the elements from the Periodic Table, we note that it does include the 
virtual-crystal parameters (average densities, valences, and core radii) 
characteristic of some of the known multi-component s,p-bonded quasicrystals, 
{\it e.g.}, Al-Mg-Li, Al(Ga)-Mg-Zn, and Al-Li(Mg)-Cu.  
However, since these ternary alloys are composed of elements with 
atomic size ratios rather far from unity, the relevance of the 
simple virtual-crystal approximation must be considered doubtful.
Clearly, the size disparity of constituent elements constitutes an 
important element in the discussion of structural stabilities of alloys.
This issue may be addressed by future extension of the DF approach 
to multi-component metallic systems, for which purpose a generalization 
of the MWDA to binary HS mixtures~\cite{DA} may prove useful.
%
%########################################################################

We now turn to the physical reasons underlying the predicted
thermodynamic stability of quasicrystalline structures.
When the effective HS packing fraction is sufficiently low that the 
free energy functional has no variational minimum, the solid exhibits
what we term a {\it phonon instability}.
Such unstable regions in the parameter space of Tables~II-IV
are readily explained by considering trends in the effective HS diameter.
In general, $d$ is seen to decrease with increasing valence 
(across a row) and with decreasing core radius (down a column), in line with 
previous remarks regarding the core region of the pair potential (see Sec.~II). 
Correspondingly, the atoms in the solid become more loosely packed and their 
density distributions broader, as reflected by the increasing Lindemann ratio 
for a given structure.  
Ultimately, when $L>0.10-0.15$ the atomic packing becomes so loose 
that the solid -- whatever its structure -- is no longer stable against 
vibrational displacements of the atoms away from their equilibrium sites 
(phonons).  
The theory is thus consistent with the Lindemann criterion for melting, 
according to which a solid becomes unstable when atomic displacements
exceed roughly $10 \%$ of the nearest-neighbor distance.
Significantly, the rational approximants, despite being less 
efficiently packed (and thus having shorter nearest-neighbor distances) 
than the crystals, have consistently smaller Lindemann ratios. 
This implies greater vibrational stiffness, which tends to enhance 
their structural stability over that of the crystals.
In passing, we note that this may be related to certain
mechanical and transport properties characteristic of many quasicrystals, 
such as high brittleness and low electrical and thermal conductivities. 
The phonon instability partly explains the appearance
of the rational approximants near the table diagonals, where the crystals 
lose mechanical stability, and the absence of any stable structure 
towards the lower-right corner of the tables.
It also accounts for the 1/1 metastability along the entire bottom row 
of Table~IV, where all other structures are mechanically unstable.

Where no entries are given in the stability tables, 
the free energy cannot be evaluated within the theory
for any of the structures considered.
The reason is revealed by closer inspection of the effective HS diameter. 
In exception to the above-noted trend, $d$ can be anomalously large
when the electron density is sufficiently high and $r_c$ sufficiently small.
Comparison with the pair potentials in Fig.~1 shows that this inflation 
of the effective hard core occurs precisely when the first minimum of 
the Friedel oscillations happens to lie beneath the core, inducing
a repulsive shoulder and pushing the first minimum of the pair potential 
out to larger distances.
[Compare, for example, the diameter in Table~II at $Z=3.2$, $r_c/r_s=0.5$ 
with its counterpart pair potential in Fig.~1 (a).]
When $d$ is so large, and $\Omega$ so small, that the maximum HS 
packing fraction 
of a given solid structure (Table~I) is exceeded, the repulsive core regions 
of nearest-neighbor atoms overlap at equilibrium separation.  Such an
energetically unfavorable configuration exhibits what we term a 
{\it packing instability}. 
In such cases, it proves impossible to find a self-consistent solution 
for the weighted density [Eq.~(\ref{rhohat2})].
The theory thus appropriately forbids such unphysical states.
Packing instability is a second reason for the 
loss of structural stability upon approaching the lower-right corner 
of Tables~II-IV, particularly at lower atomic volumes.  
Of course there is some degree of arbitrariness in splitting 
the effective pair potential at the first, rather than the second, minimum.  
We note that at $T=500$ K the characteristic thermal energy $k_BT\simeq 3$ mRy 
is comparable to the barrier height, which justifies our choice of convention. 
For shouldered potentials, however, the size of the effective hard core 
may be artificially inflated, resulting in a possible underestimate of 
quasicrystalline stability, particularly at higher densities.

The analysis of structural trends across the Periodic Table~\cite{HH}, 
as well as lattice-sum calculations~\cite{Smith,Windisch} for a wide
variety of crystal structures have shown that if the effective
pair potential develops a shoulder at a distance close to the 
nearest-neighbor separation in a close-packed structure, a lattice
distortion or finally the formation of a more open crystal structure
becomes energetically favorable.  Characteristic examples falling
within the parameter range covered by this study are the
rhombohedral structure of Hg ($Z=2$, $r_c/r_s\leq 0.40$) and
the orthorhombic structure of $\alpha$-Ga ($Z=3$, $r_c/r_s\leq 0.48$).
At even higher valence, the same effect explains the transition 
from close-packed metallic to open covalent structures.  Covalency
also entails three- and many-body forces, which 
are necessary to ensure the dynamic stability of the open structures.
Many-body forces, however, are clearly outside the scope of current 
pseudopotential and DF theories.

To this point, we have explained the loss of mechanical stability
in terms of phonon and packing instabilities.
Physical insight into the source of thermodynamic stability 
is gained by examining separately the reference and perturbation 
contributions to the total free energy. 
The reference free energy combines the entropy and that part 
of the internal energy that derives from short-range interactions, 
while the perturbation free energy is the remaining part of the 
internal energy deriving from medium- and long-range interactions.  
Figure~3 compares the reference and perturbation free energies
of the various structures as a function of $Z$ for fixed chosen 
values of $r_c/r_s$ and $\Omega$.  We emphasize that because of the 
ignored volume term (see Sec.~III) only free energy {\it differences}
are physically relevant. 
That the reference free energy consistently favors the close-packed crystals 
is to be expected, since the reference system here is a HS solid.
%#######################################################################
% Revised Text:
%
More revealing is that the perturbation free energy, 
although reinforcing the stability of the compact crystals 
at lower $Z$, increasingly disfavors them
relative to the more open quasicrystals as Z increases. 
This can be largely understood by noting that at lower $Z$, 
where the Friedel oscillations are longer in wavelength, 
the first minimum of $\phi(r)$ is relatively deep and coincides closely 
with the first nearest-neighbor distances of the highly-coordinated
fcc and hcp structures. 
With increasing $Z$, however, as the Friedel oscillations 
shorten in wavelength and move under the repulsive core, the first
minimum becomes shallower and shifts to shorter distances
more commensurate with the first nearest-neighbor distances of the 
rational approximant structures
(cf. the discussion of relative stabilities of crystalline structures 
given by Hafner and Heine (1983)~\cite{HH}).
Thus the oscillating part of the effective pair potential clearly emerges 
as the source of thermodynamic stability of simple metallic quasicrystals.
The actual order of stabilities depends on a detailed competition between 
reference and perturbation free energies or, equivalently, between 
strongly repulsive short-range and weaker medium-long-range interactions. 
%
%#######################################################################

The issue of structural stability may be framed alternatively in terms of 
a competition between internal energy $U$ and entropy $S$.  
(For clarity, it should be noted that we do not consider here the role of 
phason entropy, since construction of the rational approximants 
does not involve the breaking of any matching rules.)
Whereas in the lattice-sum method~\cite{Smith,Windisch} only $U$
is calculated, in the DF approach both $U$ and $S$ may be separated
from the free energy via the thermodynamic relations
\begin{equation}
U=F+TS
\label{U}
\end{equation}
and
\begin{equation}
S=-\left(\frac{\partial F}{\partial T}\right)_{\rho}.
\label{S}
\end{equation}
We have numerically performed the separation and individually compared 
the resulting $U$ and $S$ for the different structures.
We find in this way that entropy tends to favor the periodic crystals, 
whereas internal energy tends to favor the rational approximants.  
These tendencies, however, are not as decisive as those exhibited by 
the reference and perturbation free energies, respectively.
Thus, our DF-perturbation theory approach demonstrates the utility of 
interpreting relative structural stabilities in terms of a competition 
between short- and medium-long-range interactions, as an alternative to 
the more conventional comparison of energy and entropy.
Furthermore, it indicates that even though entropic effects
tend to favor close-packed structures, internal energy associated
with longer-range interactions may still prevail in stabilizing
quasicrystallinity.

Finally, we consider the temperature dependence of the relative stabilities.
For an effective pair potential corresponding to a point near the diagonal 
of Table~III, Fig.~4 displays the free energy per particle as a function of 
temperature, at fixed average density, for three rational approximants 
(1/1 is unstable here) and for the liquid.  
Because the zero of the vertical scale is arbitrary, only free energy 
differences are physically meaningful.
With decreasing temperature, one 
observes an inevitable liquid-solid transition.  More interesting 
is that for this case a distinct cross-over in stability 
occurs between the different approximants.  At higher temperatures, 
where short-range repulsive interactions dominate structural stability, 
the more efficiently packed short-period 2/1 structure is most stable 
(though metastable relative to the liquid).  At lower temperatures, 
medium- and long-range forces become more significant, 
ultimately stabilizing the long-period 5/3 structure. 

\section{Summary and Conclusions}

Summarizing, we have applied a density-functional-based perturbation theory 
to the question of the thermodynamic stability of one-component crystals 
and rational approximant models of quasicrystals, interacting via 
effective metallic pair potentials derived from pseudopotential theory.  
The theory is based on a splitting of the pair potential into a 
strongly repulsive short-range part and a relatively weak oscillating 
tail and the variational minimization of an approximate free energy functional 
with respect to the Gaussian-parametrized density distribution.
Comparing the free energies of several plausible solid structures
over a range of pseudopotential parameters and thermodynamic states 
corresponding to simple metals, 
we have identified general trends in relative stabilities.
With increasing electronic valence and decreasing core radius, 
the solid becomes increasingly unstable towards atomic displacements
as the atoms become more loosely packed. 
Near electronic valences and core radii for which the close-packed crystals 
lose mechanical stability, the theory predicts thermodynamic stability 
of the vibrationally stiffer quasicrystalline structures. 

Some limitations of the theory should be emphasized.
First, the DF approach, because it requires prior specification 
of the solid structure, cannot by itself predict the globally stable 
symmetry.  It can, however, predict the relative stabilities of a set 
of prescribed structures.  Although our survey includes several of the
more likely structures, it is conceivable that others not examined, 
or not treatable by the theory, may prove to be more stable.
In particular, DF theory cannot describe freezing into the more open
crystal structures ({\it e.g.}, $\alpha$,$\gamma$-Hg, $\alpha$-Ga,
$\alpha$,$\beta$-Sn) appearing close to the transition from metals
to semimetals and semiconductors.  In these materials, the liquid
is always more metallic than the solid, violating a basic assumption 
of DF theory that the interatomic potentials do not change at the transition.  
The important point, however, is that in the lower-right half of Tables~II-IV
the stability of the quasicrystalline structures will be limited by the 
higher stability of comparatively open and at least partially covalent 
structures.  We can locate the transition by drawing a line between
the points representing Hg and Ga.  This line coincides approximately
with the limits of stability of the quasicrystalline approximants
(again in good agreement with the lattice-sum calculations).  
Hence, our conclusions are not affected by this limitation.

%#####################################################################
% Modified Text:
%
Second, convergence of the perturbation expansion at first-order
[Eq. \ref{pert2}] is not ensured down to arbitrarily low temperatures.  
In principle, convergence can be established by comparing the 
first-order contribution with the residual contribution due to 
neglected higher-order terms.  
(Note that the relative magnitudes of reference and perturbation 
free energies do not necessarily reflect the degree of convergence.)
Of course in practice, the higher-order terms are unknown.
As noted in Sec. III A, however, they are proportional to fluctuations 
in the total perturbation energy and to successively higher powers of 
$1/k_BT$.  In general, if the change in $\phi_p(r)$ relative to $k_BT$ 
is sufficiently small over distances within which the single-particle 
density of the solid (or radial distribution function of the liquid)
is significant, then these terms can be safely neglected.  
For the temperatures and average densities of interest here, the solid 
density distributions (and peaks in the liquid radial distribution functions)
are so narrow that the first-order approximation is expected to be 
rather accurate.  As a typical example, consider the case of 
Fig.~1a for $Z=2.6$.  From Table II, at $T=500 K$ the stable structure 
is the fcc crystal with Lindemann ratio $L=0.101$.  The variation 
of $\phi_p(r)$ at the nearest-neighbor distance $r_{nn}\simeq 5.54 a_o$
over the relevant distance $r_{nn}L$ is considerably smaller than 
$k_BT\simeq 3$ mRy, indicating good convergence.
Caution is naturally advised, however, in extension of the 
first-order approximation to significantly lower temperatures.
%In general, the temperature range of validity will depend on 
%both the average density and the pseudopotential parameters.  
%
%#####################################################################

Third, the assumption of a single-parameter isotropic Gaussian 
density distribution is certainly of restricted validity, 
especially for very anisotropic structures.  
The close-packed crystals are known to be reasonably isotropic 
and Gaussian~\cite{YA,OLW}.
For the relatively open eight-fold coordinated bcc crystal, 
on the other hand, the isotropic assumption should not be 
particularly accurate, especially near the melting transition.
Quasicrystals and their higher-order approximants, however, are known 
to be elastically isotropic~\cite{lubensky} and also to exhibit isotropic 
dispersion relations of accoustic phonons~\cite{HK93}.
Furthermore, their relatively narrow density distributions 
(small Lindemann ratios) tend to minimize the significance 
of any deviations from Gaussian form.
Hence, an isotropic Gaussian distribution should be rather accurate 
for these structures.
In any case, the incorporation of additional degrees of freedom into a 
more flexible and realistic density parametrization would only 
result in {\it increasing} the stability of the rational approximants.  
Such technical refinements to the theory therefore should not qualitatively 
change our conclusions.

Lattice-sum studies~\cite{Smith,Windisch}, based on 
ground-state energy calculations, gave the first intriguing indication 
of {\it energetically} stable one-component metallic quasicrystals 
at zero temperature, but left open the issue of thermodynamic stability 
at finite temperatures.
Previous DF studies of hard-sphere systems~\cite{MA,Jaric2} effectively 
ruled out vibrational and configurational entropy as the source of 
quasicrystal stability, but did not consider the role of 
longer-range interactions. 
Using a practical DF-perturbation approach, we have now directly addressed 
these issues, and in the process demonstrated that the remarkable 
lattice-sum predictions carry over to finite temperatures.  
Furthermore, our approach yields new physical insight by explaining 
the stability of quasicrystals in terms of their enhanced vibrational 
stiffness and the stabilizing influence of medium- and long-range interactions.

\acknowledgements
\noindent
We thank G. Kahl for numerous helpful discussions 
and M. Kraj\^c\'i and M. Windisch 
for valuable assistance with the rational approximant structures.
This work was supported by the Fonds zur F\"orderung der wissenschaftlichen 
Forschung (Austrian Science Foundation) to whom one of us (ARD) is grateful 
for a Lise-Meitner Fellowship.

\bigskip
\noindent
*~Present address: Institut f\"ur Festk\"orperforschung, 
Forschungszentrum J\"ulich GmbH,\break 
D-52425 J\"ulich, Germany (e-mail: a.denton@kfa-juelich.de)

%%%%%%%%%%%%%%%%%%%%%%%%%%%%%%%%%%%%%%%%%%%%%%%%%%%%%%%%%%%%%%%%%%%%%%%%%%%
%%%%%%%                       Appendices
%%%%%%%%%%%%%%%%%%%%%%%%%%%%%%%%%%%%%%%%%%%%%%%%%%%%%%%%%%%%%%%%%%%%%%%%%%%

%%\appendix
%%\section{}
%%\label{app1}

%%%%%%%%%%%%%%%%%%%%%%%%%%%%%%%%%%%%%%%%%%%%%%%%%%%%%%%%%%%%%%%%%%%%%%%%%%%
%%%%%%%                       REFERENCES
%%%%%%%%%%%%%%%%%%%%%%%%%%%%%%%%%%%%%%%%%%%%%%%%%%%%%%%%%%%%%%%%%%%%%%%%%%%

%%%%%%%%%%%%%%%%%%%%%%%%%%%%%%%%%%%%%%%%%%%%%%%%%%%%%%%%%%%%%%%%%%%%%%%%%%%
%%%%%%%%                       FIGURES
%%%%%%%%%%%%%%%%%%%%%%%%%%%%%%%%%%%%%%%%%%%%%%%%%%%%%%%%%%%%%%%%%%%%%%%%%%%

%  use \protect\command{}  if you have to use a command \command{}
%  which has an argument

\newpage
\unitlength1mm
\begin{figure}
%\begin{center}
% in order to put the figures into the text you have to activate
% the line with ``\input{psfig}'' as well
%\begin{picture}(65,65)
%\put(0,0){\psfig{figure=fig1.ps,width=65mm,height=65mm}}
%\end{picture}
\noindent
\caption[]{
Effective pair potentials for several pseudopotential parameters:
(a) valence $Z=2.0$ (solid curve), $2.6$ (long-dashed), and $3.2$ 
(short-dashed) for fixed ratio of core radius to electron-sphere radius 
$r_c/r_s=0.50$ and atomic volume ${\Omega}=120 a_o^3$; 
(b) same as (a), but for ${\Omega}=150 a_o^3$; 
(c) $r_c/r_s=0.40$ (solid), $0.45$ (long-dashed), $0.50$ (short-dashed), 
and $0.55$ (dot-dashed) for fixed $Z=3.0$ and ${\Omega}=150 a_o^3$.
(See Tables~II and III for corresponding effective hard-sphere diameters 
at $T=500 K$.)
}
%\end{center}
\label{FIG1}
\end{figure}

\begin{figure}
\noindent
\caption[]{
Free energy per unit volume (in thermal units) vs. average density 
of the hard-sphere reference system:  Fluid (thick solid curve), fcc and hcp
(thick long-dashed), bcc (thick short-dashed), 1/1 (thin dot-dashed), 2/1
(thin solid), 3/2 (thin long-dashed), and 5/3 (thin short-dashed).
}
\label{FIG2}
\end{figure}

\begin{figure}
\noindent
\caption[]{
(a) Reference hard-sphere free energy, and (b) perturbation
free energy, for designated solid structures as a function of valence $Z$
at fixed core radius $r_c/r_s=0.550$, atomic volume ${\Omega}=150 a_o^3$, 
and temperature $T=500 K$.  
}
\label{FIG3}
\end{figure}

\begin{figure}
\noindent
\caption[]{
Free energy per particle vs. temperature for the liquid (solid curve)
and for the 2/1 (long-dashed), 3/2 (short-dashed), and 5/3 (dot-dashed) 
rational approximants at fixed atomic volume ${\Omega}=150 a_o^3$, 
valence $Z=3.0$, and core radius $r_c/r_s=0.500$.  
}
\label{FIG4}
\end{figure}

%%%%%%%%%%%%%%%%%%%%%%%%%%%%%%%%%%%%%%%%%%%%%%%%%%%%%%%%%%%%%%%%%%%%%%%%%%%
%%%%%%%                       TABLES
%%%%%%%%%%%%%%%%%%%%%%%%%%%%%%%%%%%%%%%%%%%%%%%%%%%%%%%%%%%%%%%%%%%%%%%%%%%

% \begin{table}
% \caption{}
% \label{}
% \end{tabular}
% \end{table}
%

\begin{table}
\noindent
\bigskip
\bigskip
\caption{
Number of basis atoms per unit cell and 
maximum packing fraction for the first four rational approximants 
to the unit-sphere packing model (see Sec.~IV A) and three elementary crystals.
}
\label{TAB1}
\begin{tabular}{ccc}
Structure&Basis Size&Maximum Packing Fraction \\
\hline
\hline
1/1&20&0.5020 \\
2/1&108&0.6400 \\
3/2&452&0.6323 \\
5/3&1904&0.6287 \\
fcc&1&0.7405 \\
hcp&2&0.7405 \\
bcc&1&0.6802 \\
\end{tabular}
\end{table}

\begin{table}
\noindent
\caption{
Thermodynamically stable structure, corresponding Lindemann ratio, 
and effective hard-sphere diameter $d$ (in Bohr radii), 
at fixed atomic volume ${\Omega}=120 a_o^3$ and temperature $T=500 K$,  
over a range of pseudopotential parameters: valence $Z$ and 
core radius $r_c$ as a ratio of electron-sphere radius $r_s$.
(Parentheses indicate metastability with respect to the liquid phase.)
}
\label{TAB2}
\begin{tabular}{cc|cccccccc}
 & & & & &$Z$& & & & \\
\hline
$r_c/r_s$& &2.0&2.2&2.4&2.6&2.8&3.0&3.2&3.4\\
\hline
\hline
0.575& &fcc&fcc&fcc&fcc&fcc&fcc&2/1&2/1\\
 & &0.00799&0.0220&0.0343&0.0441&0.0538&0.0676&0.0606&0.0763\\
 & &5.489&5.397&5.317&5.246&5.182&5.127&5.078&5.035\\
\hline
0.550& &hcp&fcc&fcc&fcc&fcc&hcp&2/1&2/1\\
 & &0.0288&0.0366&0.0468&0.0565&0.0695&0.101&0.0762&0.0941\\
 & &5.389&5.308&5.235&5.173&5.117&5.070&5.031&4.999\\
\hline
0.525& &hcp&fcc&fcc&fcc&fcc&2/1&2/1&2/1\\
 & &0.0397&0.0513&0.0610&0.0737&0.0952&0.0783&0.0934&0.107\\
 & &5.286&5.214&5.152&5.099&5.055&5.020&4.995&4.984\\
\hline
0.500& &hcp&fcc&fcc&fcc&2/1&2/1&--&--\\
 & &0.0554&0.0678&0.0804&0.101&0.0829&0.0947&--&--\\
 & &5.176&5.116&5.067&5.027&4.999&4.986&6.972&6.939\\
\hline
0.475& &hcp&fcc&fcc&2/1&--&--&--&--\\
 & &0.0767&0.0902&0.112&0.0909&--&--&--&--\\
 & &5.062&5.016&4.982&4.963&6.785&6.801&6.787&6.755\\
\hline
0.450& &hcp&fcc&2/1&--&--&--&--&--\\
 & &0.105&0.129&0.104&--&--&--&--&--\\
 & &4.944&4.915&4.905&6.581&6.618&6.621&6.601&6.568\\
\hline
0.425& &(2/1)&(2/1)&--&--&--&--&--&--\\
 & &0.111&0.127&--&--&--&--&--&--\\
 & &4.825&4.821&6.385&6.441&6.458&6.450&6.425&6.390\\
\hline
0.400& &(5/3)&--&--&--&--&--&--&--\\
 & &0.149&--&--&--&--&--&--&--\\
 & &4.709&6.196&6.273&6.304&6.305&6.287&6.257&6.220\\
\end{tabular}
\end{table}

\begin{table}
\noindent
\caption{
Same format as Table~II, but for atomic volume ${\Omega}=150 a_o^3$.
}
\label{TAB3}
\begin{tabular}{cc|cccccccc}
 & & & & &$Z$& & & & \\
\hline
$r_c/r_s$& &2.0&2.2&2.4&2.6&2.8&3.0&3.2&3.4\\
\hline
\hline
0.575& &hcp&hcp&fcc&fcc&fcc&5/3&5/3&5/3\\
 & &0.0258&0.0285&0.0395&0.0450&0.0507&0.0453&0.0526&0.0631\\
 & &5.794&5.692&5.601&5.519&5.445&5.378&5.317&5.262\\
\hline
0.550& &hcp&hcp&fcc&fcc&3/2&5/3&5/3&5/3\\
 & &0.0315&0.0356&0.0495&0.0554&0.0516&0.0581&0.0690&0.0876\\
 & &5.677&5.584&5.501&5.426&5.359&5.299&5.245&5.198\\
\hline
0.525& &hcp&fcc&fcc&fcc&5/3&5/3&(5/3)&liquid\\
 & &0.0431&0.0558&0.0620&0.0706&0.0653&0.0771&0.0967&--\\
 & &5.553&5.470&5.395&5.329&5.271&5.219&5.174&5.135\\
\hline
0.500& &hcp&fcc&fcc&(2/1)&(5/3)&(5/3)&liquid&liquid\\
 & &0.0625&0.0708&0.0802&0.0789&0.0878&0.109&--&--\\
 & &5.422&5.348&5.283&5.228&5.179&5.138&5.104&5.078\\
\hline
0.475& &hcp&fcc&(fcc)&(2/1)&(3/2)&liquid&liquid&liquid\\
 & &0.0822&0.0931&0.113&0.110&0.135&--&--&--\\
 & &5.282& 5.219&5.166&5.121&5.085&5.058&5.042&5.042\\
\hline
0.450& &(hcp)&(fcc)&(2/1)&liquid&liquid&liquid&--&--\\
 & &0.111&0.137&0.134&--&--&--&--&--\\
 & &5.136&5.085&5.044&5.013&4.993&4.989&6.964&6.955\\
\hline
0.425& &(2/1)&liquid&liquid&liquid&--&--&--&--\\
 & &0.131&--&--&--&--&--&--&--\\
 & &4.986&4.947&4.920&4.907&6.750&6.785&6.786&6.765\\
\hline
0.400& &liquid&liquid&liquid&--&--&--&--&--\\
 & &--&--&--&--&--&--&--&--\\
 & &4.831&4.806&4.796&6.569&6.619&6.629&6.614&6.584\\
\end{tabular}
\end{table}

\begin{table}
\noindent
\caption{
Same format as Table~II, but for atomic volume ${\Omega}=180 a_o^3$.
}
\label{TAB4}
\begin{tabular}{cc|cccccccc}
 & & & & &$Z$& & & & \\
\hline
$r_c/r_s$& &2.0&2.2&2.4&2.6&2.8&3.0&3.2&3.4\\
\hline
\hline
0.575& &hcp&hcp&fcc&fcc&5/3&5/3&5/3&5/3\\
 & &0.0256&0.0268&0.0426&0.0473&0.0395&0.0451&0.0520&0.0604\\
 & &6.065&5.953&5.853&5.763&5.682&5.607&5.539&5.477\\
\hline
0.550& &hcp&hcp&fcc&fcc&5/3&5/3&(5/3)&(5/3)\\
 & &0.0315&0.0398&0.0522&0.0578&0.0513&0.0582&0.0670&0.0807\\
 & &5.934&5.831&5.739&5.656&5.581&5.513&5.451&5.394\\
\hline
0.525& &hcp&fcc&fcc&5/3&5/3&(5/3)&(5/3)&(5/3)\\
 & &0.0502&0.0590&0.0647&0.0592&0.0663&0.0760&0.0919&0.129\\
 & &5.794&5.700&5.616&5.542&5.475&5.414&5.360&5.310\\
\hline
0.500& &hcp&fcc&(fcc)&(5/3)&(5/3)&(5/3)&liquid&liquid\\
 & &0.0675&0.0740&0.0830&0.0770&0.0885&0.108&--&--\\
 & &5.644&5.561&5.486&5.420&5.362&5.310&5.263&5.224\\
\hline
0.475& &hcp&(fcc)&(5/3)&(5/3)&(5/3)&liquid&liquid&liquid\\
 & &0.0865&0.0968&0.0908&0.106&0.134&--&--&--\\
 & &5.486&5.412&5.348&5.291&5.243&5.200&5.165&5.136\\
\hline
0.450& &(hcp)&(fcc)&(5/3)&liquid&liquid&liquid&liquid&liquid\\
 & &0.116&0.140&0.130&--&--&--&--&--\\
 & &5.320&5.257&5.203&5.157&5.118&5.088&5.065&5.051\\
\hline
0.425& &(2/1)&(1/1)&(1/1)&(1/1)&(1/1)&(1/1)&(1/1)&--\\
 & &0.141&0.0691&0.0758&0.0824&0.0886&0.0938&0.0970&--\\
 & &5.149&5.096&5.052&5.017&4.991&4.974&4.970&7.074\\
\hline
0.400& &(1/1)&(1/1)&(1/1)&(1/1)&(1/1)&(1/1)&--&--\\
 & &0.0853&0.0929&0.101&0.109&0.115&0.117&--&--\\
 & &4.972&4.930&4.897&4.874&4.862&4.866&6.908&6.898\\
\end{tabular}
\end{table}

\end{document}